\documentclass[fleqn,usenatbib]{mnras} 
\usepackage{graphicx}   
\usepackage{amsmath}	
\usepackage[T1]{fontenc}
\DeclareRobustCommand{\VAN}[3]{#2}
\let\VANthebibliography\thebibliography
\def\thebibliography{\DeclareRobustCommand{\VAN}[3]{##3}\VANthebibliography}

\title[Reference wavelengths of  \ion{Si}{ii}, \ion{C}{ii}, \ion{Fe}{i}, and \ion{Ni}{ii}]
      {Reference wavelengths of \ion{Si}{ii}, \ion{C}{ii}, \ion{Fe}{i}, and \ion{Ni}{ii} for quasar absorption spectroscopy}
                                                                                                      
\author[G. Nave \& C. Clear]{ Gillian Nave,$^1$
\thanks{Email:gnave@nist.gov} 
Christian Clear$^2$       \\                                                                                               
$^1$National Institute of Standards and Technology, Gaithersburg MD 20899 \\
$^2$Imperial College London, Prince Consort Road, London, SW7 2AZ, UK
}
\date{Accepted XXX. Received YYY; in original form ZZZ}
\pubyear{2020}

\begin{document}
\label{firstpage}
\pagerange{\pageref{firstpage}--\pageref{lastpage}}
\maketitle                                              

\begin{abstract}
Wavelengths of absorption lines in the spectra of galaxies along the
line-of-sight to distant quasars can be used to probe the variablility of the
fine structure constant, $\alpha$, at high redshifts, provided that the
laboratory wavelengths are known to better than 6 parts in 10$^8$, corresponding
to a radial velocity of $\approx$~20 ms$^{-1}$. For several lines of
\ion{Si}{ii}, \ion{C}{ii}, \ion{Fe}{i}, and \ion{Ni}{ii}, previously published
wavelengths are inadequate for this purpose. Improved wavelengths for these
lines were derived by re-analyzing archival Fourier transform (FT) spectra of
iron hollow cathode lamps (HCL) and a silicon carbide Penning discharge lamp,
and with new spectra of nickel HCLs. By re-optimizing the energy levels of
\ion{Fe}{i}, the absolute uncertainty of 13 resonance lines has been reduced by
over a factor of 2. A similar analysis for \ion{Si}{ii} gives improved values
for 45 lines with wavelength uncertainties over an order of magnitude smaller
than previous measurements. Improved wavelengths for 8 lines of \ion{Ni}{ii}
were measured and Ritz wavelengths from optimized energy levels determined for
an additional 3 lines at shorter wavelengths. Three lines of \ion{C}{ii} near
135~nm were observed using FT spectroscopy and the wavelengths confirm previous
measurements.
\end{abstract}


\section{Introduction}

The study of absorption lines in the spectra of galaxies along the line-of-sight
to distant quasars (quasi-stellar objects, QSO) can give important information
about the abundances, ionization and kinematics of atoms within these galaxies.
All of these studies require accurate wavelengths and oscillator strengths of
atomic lines, but the most demanding requirements for accurate wavelengths come
from the use of QSO absorption lines to study the variability of the fine
structure constant, $\alpha$, at high redshifts. In these investigations, the
wavelength separation of atomic lines in QSO absorption
spectra at different redshifts are compared to laboratory values
\citep{Dzuba_99}. Any dependence of these separations on the redshift could
suggest that $\alpha$ had a different value in the early Universe. While some
studies have found a potential variation of $\alpha$ with redshift
\citep{Murphy_03}, others have found no change \citep{Chand_06}. The
possible reported changes in $\alpha$ are less than 1:10$^5$ and require the laboratory
wavelengths to be known to a precision of better than 6 parts in 10$^8$ for
a wide variety atomic species and wavelengths. 
                                                      
\begin{figure}   
\centering       
\includegraphics[height=\columnwidth,angle=270]{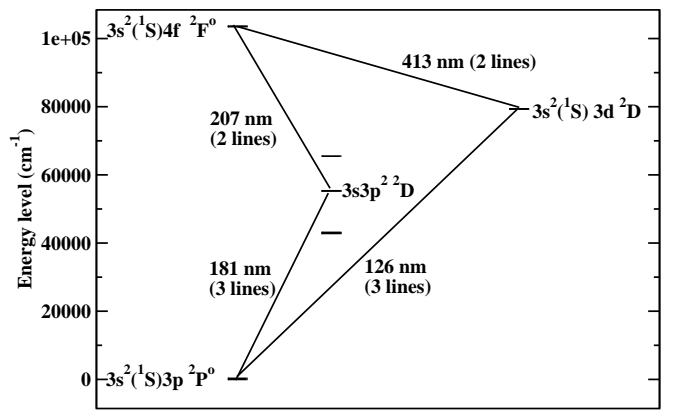} 
\caption{Partial term diagram of \ion{Si}{ii}. Measurements of spectral lines near 
181~nm, 207~nm, and 413~nm using FT spectrometry can be used to derive the 
wavelength of the lines near 126~nm.}
\label{fig:ritz}
\end{figure}

Papers by \citet{Murphy_14} and \citet{Berengut_09} include tables of spectral
lines for which the laboratory wavelength uncertainties are inadequate for the
study of QSO absorption lines, including resonance lines of \ion{Si}{ii},
\ion{C}{ii}, \ion{Fe}{i}, and \ion{Ni}{ii}. The wavelengths of some of these
lines are below 130~nm and are too short for direct measurement using Fourier
transform (FT) spectrometry. Improved wavelengths for both these lines and
weaker lines at longer wavelengths can be obtained by using Ritz wavelengths (i.e. wavelengths derived from energy level differences)
derived from energy levels optimized using FT spectrometry. For example, Fig.
\ref{fig:ritz} shows how Ritz wavelengths can be used to derive the wavelength
of the line of \ion{Si}{ii} at 126~nm, using lines at 181~nm, 207~nm and 413~nm.                                             

This paper presents improved wavelengths for resonance lines in \ion{Si}{ii},
\ion{C}{ii}, \ion{Fe}{i}, and \ion{Ni}{ii} derived from archival FT spectra 
and new measurements of nickel hollow cathode lamps (HCLs). The
spectra were used in previous studies of \ion{Fe}{i} and \ion{Fe}{ii}
\citep{Nave_94, Nave_14}, silicon and carbon ions \citep{Griesmann_00}, and a large scale analysis
of \ion{Ni}{ii} that is currently in progress \citep{Clear2018}. The
improved wavelengths were derived by assessing the calibration of the archival
spectra and optimizing the atomic energy levels to derive Ritz wavelengths.
Values for 45 lines of \ion{Si}{ii}, 13 lines of \ion{Fe}{i}, 11 lines of
\ion{Ni}{ii}, and three lines of \ion{C}{ii} are presented with uncertainties up
to an order of magnitude lower than previous values.

\section{Experimental data and analysis}                                            
\subsection{\ion{Si}{ii}}                                                                     

The last comprehensive analysis of the spectrum of \ion{Si}{ii} was by 
\citet{Shenstone_61}. \citet{Kaufman_66} measured wavelengths of 5
vacuum ultraviolet (VUV) lines with an uncertainty of 7x10$^{-5}$~nm in order to
determine the ground state splitting and the value of the $\rm 3s^24s\,^2S_{1/2}$
level. Their wavelength calibration was subsequently revised by \citet{Kaufman_74} (KE74)
in their list of reference wavelengths of atomic spectra. They combined the
\citet{Kaufman_66} measurements with additional
lines of \citet{Shenstone_61} to obtain Ritz wavelengths of 16 lines in
the VUV.
                                                   
\citet{Griesmann_00} published wavelengths of two lines of \ion{Si}{ii} measured
using FT spectroscopy of a Penning discharge source. Their spectra contain
another 30 lines of \ion{Si}{ii} that can be used to optimize the energy levels
of \ion{Si}{ii} and obtain Ritz wavelengths for lines that are too short to
measure directly using FT spectroscopy. Five of these spectra were analyzed (see
Table \ref{tab:SiC_spectra}). All of the spectra were measured using the
National Institute of Standards and Technology (NIST) FT700 FT spectrometer
\citep{Griesmann_99}. The sources were either a HCL (Spectra 1,2, and 3 in Table
\ref{tab:SiC_spectra}) or a Penning discharge lamp (Spectra 4 and 5) with SiC
cathodes run in either neon or argon. Three of the spectra (1, 2, and 3 in Table
\ref{tab:SiC_spectra}) covered the wavelength range 165~nm to 670~nm. These
spectra contained the most important lines of \ion{Si}{ii} and the Ar~II
reference wavenumbers of \citet{Whaling_95} between 434~nm and 515~nm.

Spectra recorded using FT spectroscopy are linear in wavenumber to better than
1 part in 10$^7$ and hence only one reference line is needed to put all of the 
wavenumbers on an absolute scale. In practice, many lines are usually used, 
preferably distributed throughout the spectrum in order to detect any small
non-linearities in the wavenumber scale. A multiplicative correction factor,
$k_{\mathrm{ eff}}$, is derived from these reference wavenumbers, 
$\sigma_{\mathrm{ref}}$, and the observed wavenumbers, $\sigma_{\mathrm{obs}}$, 
using:
\begin{equation}
k_{\mathrm{eff}} = \sigma_{\mathrm{ref}}/\sigma_{\mathrm{obs}}-1
\end{equation}
This is then applied to the observed wavenumbers to obtain the corrected wavenumbers:
\begin{equation}
\sigma_{\mathrm{corr}} = (1 + k_{\mathrm{eff}})\sigma_{\mathrm{obs}}
\end{equation}

\begin{figure} 
\centering     
\includegraphics[height=\columnwidth,angle=270]{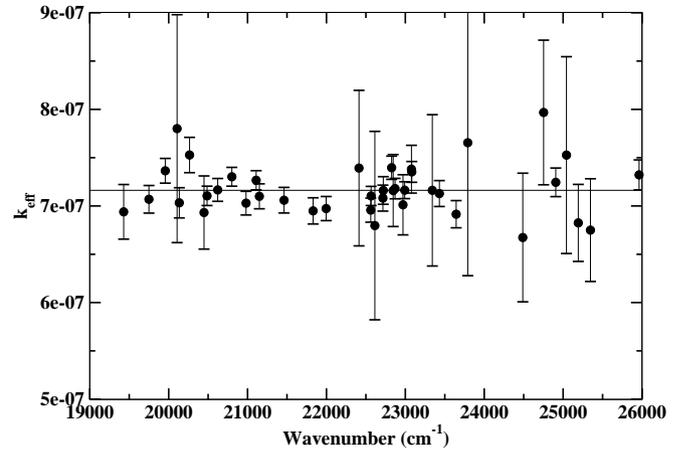}   
\caption{Wavelength correction factor $k_{\text{eff}}$ derived using Ar~II lines 
observed in spectrum 3 in Table \ref{tab:SiC_spectra}. The horizontal line is
the weighted average value of $k_{\text{eff}}$ of 7.16$\times$10$^{-7}$. }                                                      
\label{fig:SiC_wavecal}
\end{figure}

The wavenumbers in spectra 1, 2, and 3 in Table \ref{tab:SiC_spectra} were
calibrated directly from the Ar~II reference wavenumbers of \cite{Whaling_95}
and an example calibration is shown in Fig. \ref{fig:SiC_wavecal}. Since all of
the calibration lines are at wavenumbers below 26500~cm$^{-1}$ (wavelengths above 377~nm), the
calibration was confirmed with Si~II lines at larger wavenumbers
\citep{Griesmann_00} to place an upper limit on any potential small slope in the
calibration. The Si~II lines agreed within 4 parts in 10$^{8}$  and this value
was adopted as a minimum calibration uncertainty for all the spectra in Table
\ref{tab:SiC_spectra}.

\begin{table*}  
\caption{Spectra of SiC Penning discharge used for Ritz wavelengths in
\ion{Si}{ii}.}
\centering                                                                                    
\begin{tabular}{lllllllll}
\hline
Ref.& Spectrum name  & Spectral range  & Gas   & Pressure & Current & $k_{\text{eff}}$         & Detector & Filter \\      
No. &                &    (cm$^{-1}$)  &       &   (Pa)   &   (A)   &                \\          
\hline                                                                                           
1   & sic-290999.001 & 15~000 - 60~000 & Ar,Ne & 130      & 0.6     & (8.65$\pm$0.03)x10$^{-7}$  &  R106UH  & none \\   
2   & sic-290999.002 & 15~000 - 60~000 & Ar,Ne & 140      & 0.7     & (7.23$\pm$0.03)x10$^{-7}$  &  R106UH  & none \\
3   & sic-290999.003 & 15~000 - 60~000 & Ar,Ne & 140      & 0.7     & (7.16$\pm$0.03)x10$^{-7}$  &  R106UH  & none \\  
4   & sic-030999.012 & 52~000 - 77~000 & Ne    & 1.2      & 2.0     & (18.5$\pm$0.05)x10$^{-7}$  &  R1259   & 160~nm \\
5   & sic-030999.014 & 52~000 - 77~000 & Ne    & 0.2      & 1.5     & (19.0$\pm$0.05)x10$^{-7}$  &  R1259   & 160~nm \\
\hline                                                                  
\end{tabular}     
\label{tab:SiC_spectra}     
\end{table*}
    
\begin{figure} 
\centering     
\includegraphics[height=\columnwidth,angle=270]{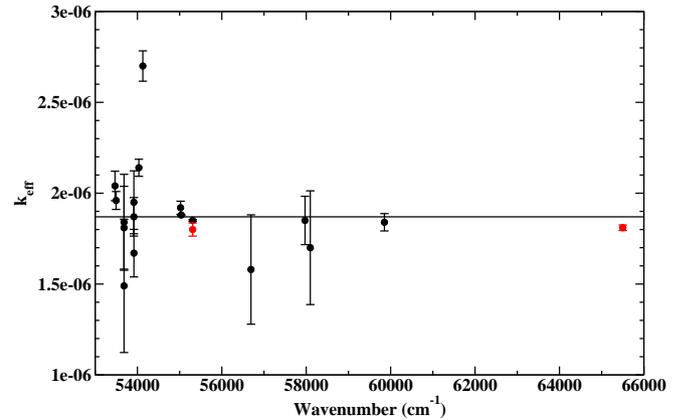}   
\caption{Wavelength correction factor k$_{\text{eff}}$ derived using Ar~II lines 
observed in spectrum 4 in Table \ref{tab:SiC_spectra}. The black points are derived from
overlapping wavenumber standards taken from spectrum 1. The red points are derived from
Si~II wavenumber taken from \citet{Griesmann_00}. The horizontal line is the weighted
average value of $k_{\text{eff}}$ of 1.85$\times$10$^{-6}$. }                                                      
\label{fig:SiC_4_wavecal}
\end{figure}

Spectra 4 \& 5 in Table \ref{tab:SiC_spectra} cover the                       
wavelength region 130~nm to 192~nm. The wavenumbers in these spectra were                                 
calibrated using lines measured in the other three spectra and Ritz wavenumbers                         
of \ion{Si}{ii} determined from lines measured in spectra 1, 2, and 3 in Table
\ref{tab:SiC_spectra}. An example calibration plot for spectrum 4 is shown in
Fig. \ref{fig:SiC_4_wavecal}.

Optimized energy levels and Ritz wavelengths in \ion{Si}{ii} were derived using
the {\sc lopt} computer program of \citet{Kramida_11}. Weights were assigned as
the squared reciprocal of the uncertainty. This uncertainty was estimated by
combining in quadrature the statistical measurement of the line's position with
the calibration uncertainty. The statistical uncertainty $\delta_{stat}$ was
estimated from the full width at half maximum, $\mathit{FWHM}$, the resolution, $r$, and
the signal-to-noise ratio, $S/N$, using: 
\begin{equation}\label{eq:unc_line}
\delta_{stat} = \frac{\sqrt{\mathit{({FWHM})}\times{\mathit{r}}}}{S/N}
\end{equation}                                        
This was derived from equation 9.2 of \citet{Davis_01} with the geometrical factor, $f = 1$, 
and the number of independent points, $N_w = {\mathit{FWHM}}/r$ . The weighted average and
uncertainty of all measurements was taken, as described in \cite{Liggins_2020}.

The level optimization was carried out in two steps. First, the values of 22
energy levels were derived using 32 observed lines in the SiC Penning discharge.
Ritz wavelengths were obtained for these lines and an additional 7 lines to the
two levels of the ground term, $3s^2(^1S)3p\,^2\!P^{\circ}$. The FT spectra in
Table \ref{tab:SiC_spectra} do not include any lines that can be used to derive
the values of the $3s3p^2\,^2\!S_{1/2}$ or $3s3p^2\, ^2\!P_{1/2,3/2}$ levels as
all the strong lines from these levels lie below the wavelength limit for FT
spectroscopy. However, improved values for these levels can be derived by
including the grating wavelengths of KE74. By comparing the Ritz
wavelengths from {\sc lopt} with those of KE74, it was determined
that their wavelengths should be reduced by $(5\pm 2)\times10^{-5}$~nm in order
to put them on the same wavelength scale as the FT spectra. Wavelengths for 4
lines near 119~nm and two lines near 130~nm were thus taken from
KE74, reduced by $5\times 10^{-5}$~nm, and included with the 32
lines from the FT spectra in a second level optimization. This determined the
values of 25 levels in \ion{Si}{ii} and Ritz wavelengths for 45 lines. The Ritz
wavelengths are given in Table \ref{tab:SiII_ritz}, with the corresponding
energy levels in Table \ref{tab:SiII_lev}.                                     
                                                      
\begin{table*}    
\caption{Ritz wavelengths and wavenumbers of lines of \ion{Si}{ii}.}  
\begin{tabular}{llllllll}
\hline
Ritz air   & Ritz vacuum      &  Ritz            & Observed           & Lower                     & Upper                      & Previous          & Ref.$^c$ \\
wavelength & wavelength$^a$   &  wavenumber$^a$  & wavenumber$^{a,b}$ & level                     & level                      & wavelength        &      \\
(nm)       & (nm)             &  (cm$^{-1}$)     & (cm$^{-1}$)        &                           &                            &   (nm)            &      \\
 \hline                                                                                                   
     -     &  98.987283  (3)  & 101023.078  (3)  &     -              & $\rm 3s^2 3p\, ^2P^{\circ}_{1/2}$ & $\rm 3s^2 4d \, ^2D_{3/2}$ & 98.98730 (10)     & (1) \\
     -     &  99.2682467 (20) & 100737.1474 (21) &     -              & $\rm 3s^2 3p\, ^2P^{\circ}_{3/2}$ & $\rm 3s^2 4d \, ^2D_{5/2}$ & 99.26826 (10)     & (1) \\          
     -     & 102.069850  (10) &  97972.124  (9)  &     -              & $\rm 3s^2 3p\, ^2P^{\circ}_{1/2}$ & $\rm 3s^2 5s \, ^2S_{1/2}$ & 102.06990 (10)    & (2) \\
     -     & 102.369975  (10) &  97684.893  (9)  &     -              & $\rm 3s^2 3p\, ^2P^{\circ}_{3/2}$ & $\rm 3s^2 5s \, ^2S_{1/2}$ & 102.37003 (10)    & (2) \\
     -     & 119.041555 (14)  &  84004.279  (10) & 84004.283$^d$ (14) & $\rm 3s^2 3p\, ^2P^{\circ}_{1/2}$ & $\rm 3s 3p^2 \, ^2P_{3/2}$ & 119.04160 (10)    & (2) \\
     -     & 119.328941 (14)  &  83801.967  (10) & 83801.975$^d$ (14) & $\rm 3s^2 3p\, ^2P^{\circ}_{1/2}$ & $\rm 3s 3p^2 \, ^2P_{1/2}$ & 119.32898 (10)    & (2) \\
     -     & 119.449984 (14)  &  83717.048  (10) & 83717.043$^d$ (14) & $\rm 3s^2 3p\, ^2P^{\circ}_{3/2}$ & $\rm 3s 3p^2 \, ^2P_{3/2}$ & 119.45004 (10)    & (2) \\
     -     & 119.739348 (14)  &  83514.736  (10) & 83514.727$^d$ (14) & $\rm 3s^2 3p\, ^2P^{\circ}_{3/2}$ & $\rm 3s 3p^2 \, ^2P_{1/2}$ & 119.73941 (10)    & (2) \\                 
     -     & 126.042174  (7)  &  79338.524  (4)  &     -              & $\rm 3s^2 3p\, ^2P^{\circ}_{1/2}$ & $\rm 3s^2 3d \, ^2D_{3/2}$ & 126.04223 (10)    & (2) \\   
     -     & 126.473770  (5)  &  79067.778  (3)  &     -              & $\rm 3s^2 3p\, ^2P^{\circ}_{3/2}$ & $\rm 3s^2 3d \, ^2D_{5/2}$ & 126.47379 (10)    & (2) \\
     -     & 126.500145  (6)  &  79051.293  (4)  &     -              & $\rm 3s^2 3p\, ^2P^{\circ}_{3/2}$ & $\rm 3s^2 3d \, ^2D_{3/2}$ & 126.50022 (10)    & (2) \\
     -     & 130.437070 (15)  &  76665.322  (9)  & 76665.328$^d$ (12) & $\rm 3s^2 3p\, ^2P^{\circ}_{1/2}$ & $\rm 3s 3p^2 \, ^2S_{1/2}$ & 130.43711 (10)    & (2) \\                        
     -     & 130.927598 (15)  &  76378.091  (9)  & 76378.084$^d$ (12) & $\rm 3s^2 3p\, ^2P^{\circ}_{3/2}$ & $\rm 3s 3p^2 \, ^2S_{1/2}$ & 130.92766 (10)    & (2) \\
     -     & 152.670698  (6)  &  65500.454  (3)  & 65500.456$^e$ (5)  & $\rm 3s^2 3p\, ^2P^{\circ}_{1/2}$ & $\rm 3s^2 4s \, ^2S_{1/2}$ & 152.670698 (2)$^f$& (3) \\
     -     & 153.343135  (5)  &  65213.2228 (22) & 65213.226$^e$ (5)  & $\rm 3s^2 3p\, ^2P^{\circ}_{3/2}$ & $\rm 3s^2 4s \, ^2S_{1/2}$ & 153.34318 (10)    & (2) \\
     -     & 171.083389  (22) &  58451.028  (8)  & 58451.027  (9)     & $\rm 3s 3p^2\, ^2D_{3/2}$ & $\rm 3s^2 5f \, ^2F^{\circ}_{5/2}$ & 171.0826          & (4) \\
     -     & 171.130279  (15) &  58435.013  (5)  & 58435.012  (6)     & $\rm 3s 3p^2\, ^2D_{5/2}$ & $\rm 3s^2 5f \, ^2F^{\circ}_{7/2}$ & 171.1296          & (4) \\
     -     & 180.801278  (7)  &  55309.3437 (23) & 55309.342  (3)     & $\rm 3s^2 3p\, ^2P^{\circ}_{1/2}$ & $\rm 3s 3p^2 \, ^2D_{3/2}$ & 180.801288 (1)$^f$& (3) \\ 
     -     & 181.692840  (6)  &  55037.9422 (19) & 55037.942  (3)     & $\rm 3s^2 3p\, ^2P^{\circ}_{3/2}$ & $\rm 3s 3p^2 \, ^2D_{5/2}$ & 181.69290 (10)    & (2) \\
     -     & 181.745112  (6)  &  55022.1126 (19) & 55022.112  (3)     & $\rm 3s^2 3p\, ^2P^{\circ}_{3/2}$ & $\rm 3s 3p^2 \, ^2D_{3/2}$ & 181.74517 (10)    & (2) \\
205.86445  & 205.93032   (3)  &  48560.116  (6)  & 48560.119  (9)     & $\rm 3s 3p^2\, ^2D_{5/2}$ & $\rm 3s^2 5p \, ^2P^{\circ}_{3/2}$ & 205.8646          & (4) \\
205.90117  & 205.96704   (5)  &  48551.457  (11) & 48551.476  (18)    & $\rm 3s 3p^2\, ^2D_{3/2}$ & $\rm 3s^2 5p \, ^2P^{\circ}_{1/2}$ & 205.9014          & (4) \\
207.201374 & 207.267492  (15) &  48246.833  (3)  & 48246.832  (4)     & $\rm 3s 3p^2\, ^2D_{3/2}$ & $\rm 3s^2 4f \, ^2F^{\circ}_{5/2}$ & 207.2016          & (4) \\
207.269758 & 207.335890  (12) &  48230.917  (3)  & 48230.917  (3)     & $\rm 3s 3p^2\, ^2D_{5/2}$ & $\rm 3s^2 4f \, ^2F^{\circ}_{7/2}$ & 207.2701          & (4) \\                    
233.44080  & 233.51239   (9)  &  42824.280  (16) & 42824.272  (22)    & $\rm 3s^2 3p\, ^2P^{\circ}_{1/2}$ & $\rm 3s 3p^2 \, ^4P_{1/2}$ & 233.4404          & (4) \\
233.46084  & 233.53244   (10) &  42820.604  (18) & 42820.604  (18)    & $\rm 3s^2 3p\, ^2P^{\circ}_{3/2}$ & $\rm 3s 3p^2 \, ^4P_{5/2}$ & 233.4606          & (4) \\ 
234.42046  & 234.49227   (23) &  42645.33   (4)  & 42645.33   (4)     & $\rm 3s^2 3p\, ^2P^{\circ}_{3/2}$ & $\rm 3s 3p^2 \, ^4P_{3/2}$ & 234.4203          & (4) \\
235.01724  & 235.08918   (9)  &  42537.049  (16) & 42537.060  (25)    & $\rm 3s^2 3p\, ^2P^{\circ}_{3/2}$ & $\rm 3s 3p^2 \, ^4P_{1/2}$ & 235.0174          & (4) \\
290.42808  & 290.51316   (7)  &  34421.848  (8)  & 34421.852  (13)    & $\rm 3s^2 3d\, ^2D_{3/2}$ & $\rm 3s^2 5f \, ^2F^{\circ}_{5/2}$ & 290.4283          & (4) \\
290.56881  & 290.65393   (5)  &  34405.177  (5)  & 34405.180  (9)     & $\rm 3s^2 3d\, ^2D_{5/2}$ & $\rm 3s^2 5f \, ^2F^{\circ}_{7/2}$ & 290.5692          & (4) \\
333.31364  & 333.40951   (14) &  29993.146  (13) & 29993.151  (23)    & $\rm 3s^2 4p\, ^2P^{\circ}_{1/2}$ & $\rm 3s^2 6s \, ^2S_{1/2}$ & 333.3139          & (4) \\
333.98154  & 334.07758   (14) &  29933.167  (13) & 29933.165  (15)    & $\rm 3s^2 4p\, ^2P^{\circ}_{3/2}$ & $\rm 3s^2 6s \, ^2S_{1/2}$ & 333.9819          & (4) \\
385.366036 & 385.475318  (19) &  25941.9982 (13) & 25941.9980 (16)    & $\rm 3s 3p^2\, ^2D_{3/2}$ & $\rm 3s^2 4p \, ^2P^{\circ}_{3/2}$ & 385.3664          & (4) \\
385.601333 & 385.710675  (16) &  25926.1686 (11) & 25926.1686 (11)    & $\rm 3s 3p^2\, ^2D_{5/2}$ & $\rm 3s^2 4p \, ^2P^{\circ}_{3/2}$ & 385.6017          & (4) \\
386.259098 & 386.368612  (16) &  25882.0196 (10) & 25882.0194 (11)    & $\rm 3s 3p^2\, ^2D_{3/2}$ & $\rm 3s^2 4p \, ^2P^{\circ}_{1/2}$ & 386.2595          & (4) \\
407.54437  & 407.65943   (10) &  24530.280  (6)  & 24530.277  (7)     & $\rm 3s^2 3d\, ^2D_{5/2}$ & $\rm 3s^2 5p \, ^2P^{\circ}_{3/2}$ & 407.5451          & (4) \\
407.67737  & 407.79247   (18) &  24522.277  (11) & 24522.266  (13)    & $\rm 3s^2 3d\, ^2D_{3/2}$ & $\rm 3s^2 5p \, ^2P^{\circ}_{1/2}$ & 407.6781          & (4) \\
412.805492 & 412.921940  (19) &  24217.6524 (11) & 24217.6524 (11)    & $\rm 3s^2 3d\, ^2D_{3/2}$ & $\rm 3s^2 4f \, ^2F^{\circ}_{5/2}$ & 412.8067          & (4) \\
413.088166 & 413.204688  (19) &  24201.0807 (11) & 24201.0807 (11)    & $\rm 3s^2 3d\, ^2D_{5/2}$ & $\rm 3s^2 4f \, ^2F^{\circ}_{7/2}$ & 413.0893          & (4) \\
504.102259 & 504.242828  (24) &  19831.7149 (10) & 19831.7147 (10)    & $\rm 3s^2 4p\, ^2P^{\circ}_{1/2}$ & $\rm 3s^2 4d \, ^2D_{3/2}$ & 504.1026          & (4) \\
505.598248 & 505.739215  (23) &  19773.0366 (9)  & 19773.0366 (9)     & $\rm 3s^2 4p\, ^2P^{\circ}_{3/2}$ & $\rm 3s^2 4d \, ^2D_{5/2}$ & 505.5981          & (4) \\
505.63150  & 505.77247   (4)  &  19771.7363 (17) & 19771.738  (3)     & $\rm 3s^2 4p\, ^2P^{\circ}_{3/2}$ & $\rm 3s^2 4d \, ^2D_{3/2}$ & 505.6314          & (4) \\ 
595.7555   & 595.9205    (3)  &  16780.761  (9)  & 16780.765  (14)    & $\rm 3s^2 4p\, ^2P^{\circ}_{1/2}$ & $\rm 3s^2 5s \, ^2S_{1/2}$ & 595.7561          & (4) \\ 
597.8925   & 598.0581    (3)  &  16720.782  (9)  & 16720.780  (12)    & $\rm 3s^2 4p\, ^2P^{\circ}_{3/2}$ & $\rm 3s^2 5s \, ^2S_{1/2}$ & 597.8929          & (4) \\ 
634.70935  & 634.88484   (6)  &  15750.8880 (14) & 15750.8886 (15)    & $\rm 3s^2 4s\, ^2S_{1/2}$ & $\rm 3s^2 4p \, ^2P^{\circ}_{3/2}$ & 634.7103          & (4) \\ 
\hline                                                                   
\end{tabular}                                     

{\footnotesize
$^a$ One standard uncertainty in the last digits given in parenthesis.                                                       
$^b$ Observed wavenumber is weighted average from spectra 1,2 \& 3 in Table \ref{tab:SiC_spectra} unless marked.
$^c$ (1)\citet{Kaufman_66}; (2) \citet{Kaufman_74} (KE74); (3) \citet{Griesmann_00}; (4) \citet{Shenstone_61}.         
$^d$ Observed wavenumber derived from the wavelength from KE74, reduced by 5$\times$10$^{-5}$~nm.                    
$^e$ Observed wavenumber taken from spectrum 4 in Table \ref{tab:SiC_spectra}.                                  
$^f$ See section \ref{Si_comp} for discussion of this uncertainty.                                              
}
\label{tab:SiII_ritz} 
\end{table*}                                                                                                     

\begin{table}   
\caption{Energy levels of \ion{Si}{ii}}                          
\begin{tabular}{lllll}   
\hline       
Configuration & Term        & Level value & Unc$^a$     & Number of  \\
              &             &             &             & transitions \\                               
              &             & (cm$^{-1}$) & (cm$^{-1}$) &         \\
\hline                                    
$\rm 3s^2 3p$ & $^2P^{\circ}_{1/2}$ &      0.0000 &             & 3     \\    
$\rm 3s^2 3p$ & $^2P^{\circ}_{3/2}$ &    287.231  &  0.003      & 6     \\      
$\rm 3s 3p^2$ & $^4P_{1/2}$         &  42824.280  &  0.016      & 2     \\      
$\rm 3s 3p^2$ & $^4P_{3/2}$         &  42932.56   &  0.04       & 1     \\      
$\rm 3s 3p^2$ & $^4P_{5/2}$         &  43107.835  &  0.018      & 1     \\      
$\rm 3s 3p^2$ & $^2D_{3/2}$         &  55309.3437 &  0.0023     & 7     \\                     
$\rm 3s 3p^2$ & $^2D_{5/2}$         &  55325.173  &  0.003      & 5     \\      
$\rm 3s^2 4s$ & $^2S_{1/2}$         &  65500.454  &  0.003      & 3     \\                   
$\rm 3s 3p^2$ & $^2S_{1/2}$         &  76665.322  &  0.009      & 2$^b$ \\
$\rm 3s^2 3d$ & $^2D_{3/2}$         &  79338.524  &  0.004      & 3     \\      
$\rm 3s^2 3d$ & $^2D_{5/2}$         &  79355.009  &  0.004      & 3     \\                        
$\rm 3s^2 4p$ & $^2P^{\circ}_{1/2}$ &  81191.3633 &  0.0025     & 4     \\      
$\rm 3s^2 4p$ & $^2P^{\circ}_{3/2}$ &  81251.3419 &  0.0025     & 7     \\      
$\rm 3s 3p^2$ & $^2P_{1/2}$         &  83801.967  &  0.010      & 2$^b$ \\
$\rm 3s 3p^2$ & $^2P_{3/2}$         &  84004.279  &  0.010      & 2$^b$ \\
$\rm 3s^2 5s$ & $^2S_{1/2}$         &  97972.124  &  0.009      & 2     \\                                   
$\rm 3s^2 4d$ & $^2D_{3/2}$         & 101023.078  &  0.003      & 2     \\                     
$\rm 3s^2 4d$ & $^2D_{5/2}$         & 101024.379  &  0.003      & 1     \\      
$\rm 3s^2 4f$ & $^2F^{\circ}_{7/2}$ & 103556.090  &  0.004      & 2     \\      
$\rm 3s^2 4f$ & $^2F^{\circ}_{5/2}$ & 103556.176  &  0.004      & 2     \\                                                       
$\rm 3s^2 5p$ & $^2P^{\circ}_{1/2}$ & 103860.801  &  0.011      & 2     \\      
$\rm 3s^2 5p$ & $^2P^{\circ}_{3/2}$ & 103885.289  &  0.007      & 2     \\      
$\rm 3s^2 6s$ & $^2S_{1/2}$         & 111184.509  &  0.013      & 2     \\      
$\rm 3s^2 5f$ & $^2F^{\circ}_{7/2}$ & 113760.186  &  0.006      & 2     \\      
$\rm 3s^2 5f$ & $^2F^{\circ}_{5/2}$ & 113760.372  &  0.008      & 2     \\                             
\hline                                                               
\end{tabular}    

{\footnotesize
$^a$ One standard uncertainty of level value with respect to the ground level.
$^b$ Level determined by lines taken from KE74
}        
\label{tab:SiII_lev} 
\end{table}

\subsubsection{Comparison of \ion{Si}{ii} wavelengths with previous results}\label{Si_comp}                                                         

Two of the lines in Table \ref{tab:SiII_ritz} are also reported by
\citet{Griesmann_00}, who used some of the same spectra as this work. The
wavelength of the line at 152.67~nm is the same as their value, but the observed
wavelength of the line at 180.80~nm differs by $1\times10^{-5}$~nm, or 6 parts
in 10$^8$. This discrepancy is 1.3 times the joint uncertainties of the values
in Table \ref{tab:SiII_ritz} and is likely due to the use of a different set of
spectra. Although the wavelengths reported by \citet{Griesmann_00} have
uncertainties as small as 5 parts in 10$^9$, this uncertainty is smaller than
that of the Ar~II wavenumber standards that they used to calibrate their
wavenumber scale. Their uncertainties are thus likely to be underestimated.
Their calibration uncertainty also does not account for any possible error in
the transfer of the wavenumber calibration from the region of the Ar~II lines
near 450~nm down to the VUV. We regard the uncertainties of both the observed
and Ritz wavelengths in Table \ref{tab:SiII_ritz} to be a more realistic
estimate.

Ten of the 16 \ion{Si}{ii} wavelengths in KE74 have new Ritz values derived from
our FT spectra in Table \ref{tab:SiII_ritz}, including the two lines near
152.67~nm and 180.80~nm reported by \citet{Griesmann_00}. \citet{Reader_02}
compared the wavelengths of \citet{Griesmann_00} with those of KE74 for these
two lines, finding that the two wavelength scales agree to $1\times10^{-5}$~nm,
well within the quoted uncertainty of $5\times10^{-5}$~nm of KE74. Comparison of
all ten \ion{Si}{ii} lines shows that the wavelengths of KE74 are $(5\pm
2)\times10^{-5}$~nm larger than the Ritz wavelengths derived from our FT
spectra. This value was thus used to adjust the scale of the remaining six lines
of KE74, and these lines were then incorporated into the final level
optimization.  The wavelength of the \ion{Si}{ii} forbidden line at
34.8152~$\mu$m was measured by \citet{Feuchtgruber_97} in planetary nebulae with
the Short Wavelength Spectrometer on the Infrared Space Observatory. Their value
corresponds to a fine structure interval in the ground term of
287.231$\pm$0.005~cm$^{-1}$, the same as the value in Table \ref{tab:SiII_lev},
but with a slightly larger uncertainty.

\subsection{\ion{C}{ii}}   

\begin{figure*}    
\centering   
\includegraphics[height=2\columnwidth,angle=270]{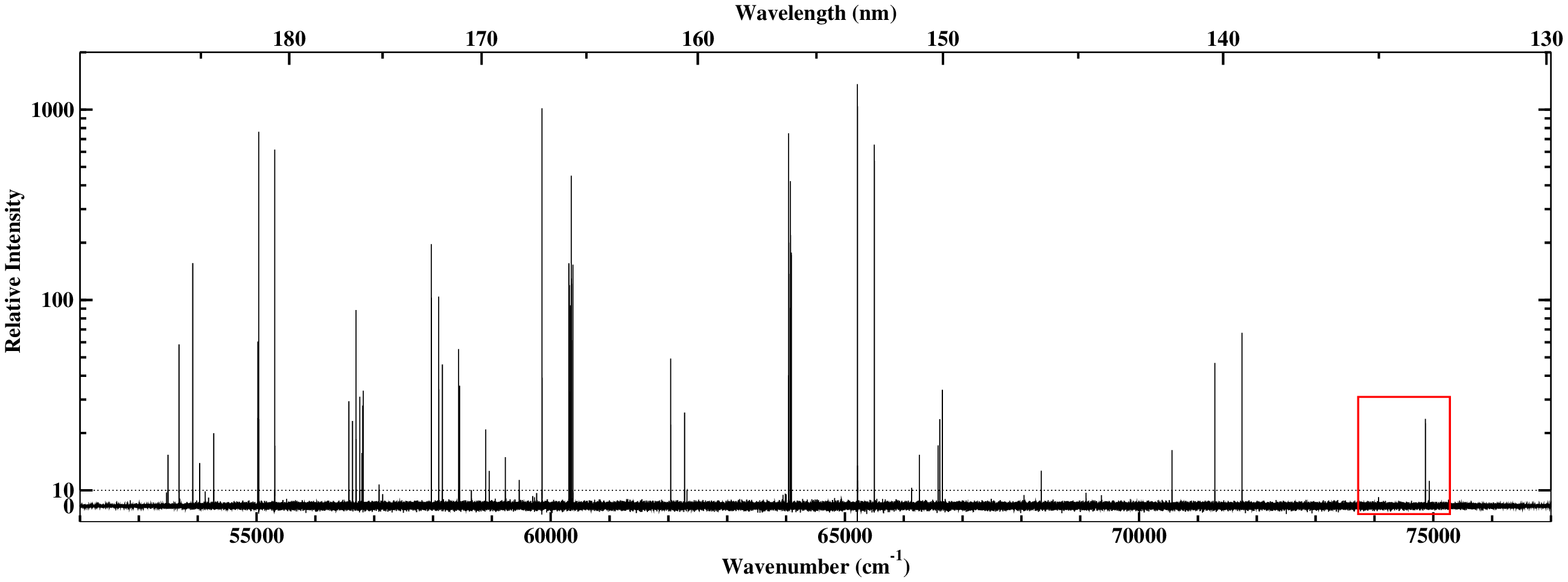}   
\caption{Spectrum of Penning discharge lamp with SiC cathodes (Spectrum 4 in 
Table \ref{tab:SiC_spectra}). The plot is linear for relative intensities 
below 10 and logarithmic for relative intensities above 10. The red box outlines
the position of the three C II lines near 133~nm}
\label{fig:SiC_plot}
\end{figure*}

One of the spectra of the SiC Penning discharge included three strong lines near
133~nm due to \ion{C}{ii} that had not been reported by \citet{Griesmann_00}
(see Fig. \ref{fig:SiC_plot}). This wavelength is shorter than previously
reported using a transmissive beamsplitter in a FT spectrometer
\citep{Thorne_96} and approaches the  $\approx$~125~nm cut-off wavelength of the calcium fluoride
beamsplitter in our instrument. These lines are probably enhanced in this
spectrum by two factors. First, the spectrum of the SiC Penning discharge is
relatively sparse, with less than 100 lines in the region from 130~nm to 187~nm.
The spectral filling factor in equation 2 of \citet{Thorne_96} is thus small
compared to typical iron-group spectra containing several hundred lines in this
range. Second, the noise from spectral lines outside the region of interest was
eliminated by using a photomultiplier with a quantum efficiency of over 20~\% at
135~nm and almost no response above 180~nm. The region was narrowed further by
using a metal-dielectric reflection filter centered at 160~nm. The wavelengths
and wavenumbers of the \ion{C}{ii} lines are given in Table \ref{tab:CII}. The
uncertainties are dominated by the statistical uncertainty of the measurement of
the position of the line. This uncertainty is about 3 parts in 10$^6$ for the
$2s^22p\,^2\!P^{\circ}_{3/2} - 2s2p^2\,^2\!D_{3/2} $ transition. An improved
wavelength for this line can be determined by adding the wavenumber of the
$2s^22p\,^2\!P^{\circ}_{1/2} - 2s2p^2\,^2\!D_{3/2} $ transition to the fine
structure splitting of the ground term in \ion{C}{ii} from \citet{Cooksy_86}.
The values agree with the measurements of \citet{Herzberg_58} within the joint
uncertainties.
   
\begin{table*}    
\caption{Wavelengths and wavenumbers for lines of \ion{C}{ii}.}  
\begin{tabular}{llllllllll}                                            
\hline
Observed        & Observed          & Lower                      & Upper                    & Previous   $^{a,b}$ \\
wavenumber$^a$  & wavelength$^a$    & level                      & level                    & wavelength      \\                                                                      
(cm$^{-1}$)     &      (nm)         &                            &                          &   (nm)          \\
\hline                                                                                                       
74866.651 (20)  &  133.57082 (4)    & $\rm 2s^22p\,^2P^{\circ}_{3/2} $ & $\rm 2s2p^2\,^2D_{5/2} $ & 133.57077 (7)   \\ 
74869.20$^c$(4) &  133.56626$^c$(8) & $\rm 2s^22p\,^2P^{\circ}_{3/2} $ & $\rm 2s2p^2\,^2D_{3/2} $ & 133.56625 (7)   \\
74932.60  (4)   &  133.45326 (8)    & $\rm 2s^22p\,^2P^{\circ}_{1/2} $ & $\rm 2s2p^2\,^2D_{3/2} $ & 133.45323 (7)   \\
\hline
\end{tabular}                                     
  
{\footnotesize   
$^a$ One standard uncertainty in the last digits given in parenthesis.  
$^b$ Previous wavelengths are from \citet{Herzberg_58}.             
$^c$ Wavenumber and wavelength derived from line at 133.45 nm and the
$2p\,^2\!P^{\circ}_{1/2} - 2p\,^2\!P^{\circ}_{3/2}$ ground term splitting of 
63.3951 cm$^{-1}$ from \citet{Cooksy_86}. The observed wavenumber is 
74869.31~(23)~cm$^{-1}$, corresponding to a wavelength of 133.5661 (4) nm. \\       
}
\label{tab:CII}                  
\end{table*}                                              
   
\subsection{\ion{F{e}}{i}}
        
The wavenumber calibration of the \ion{Fe}{i} lines reported by
\citet{Nave_94} has been revised by  \citet{Nave_11}, who recommended an 
increase in the wavenumbers in these papers of 10.6 parts in
10$^8$ in the ultraviolet and 6.7 parts in 10$^8$ in the visible and infrared.
The measured wavenumbers in \citet{Nave_94} are given to 3 decimal
places, with a conservative estimated uncertainty of 0.005 cm$^{-1}$ for the
strong lines. More accurate Ritz wavenumbers for the resonance lines in \ion{Fe}{i} can
be obtained from the original data by adjusting the wavenumber scale and
re-optimizing the energy levels with improved techniques using the {\sc lopt} 
computer code \citep{Kramida_11}.

New values for 829 levels in \ion{Fe}{i} were derived from the wavenumbers of 9349
lines using the {\sc LOPT} computer code of \citet{Kramida_11}. The weights and
uncertainties used in the optimization were estimated in a similar way to the
\ion{Si}{ii} levels and full details are given in \citet{Nave_11}. For the levels giving
the strongest resonance lines observed in QSO absorption spectra, the
uncertainty of the levels is dominated by the calibration uncertainty of
4:10$^8$. Table \ref{tab:FeI} presents the Ritz wavenumbers and uncertainties of
13 lines of most interest for QSO absorption line spectra  \citep{Murphy_14}. The wavenumbers are
1.06 parts in 10$^7$ larger than those from \citet{Nave_91} and \citet{Nave_94},
as noted in \citet{Nave_11}, and are the same as given in the NIST Atomic Spectra
Database \citep{Ralchenko}, but with reduced uncertainties.                                      

\begin{table*}       
\caption{Ritz wavelengths and wavenumbers of resonance lines in \ion{Fe}{i}}  
\begin{tabular}{llllllllll}    
\hline                                                                                                                               
Ritz Air $^a$ & Ritz vacuum $^a$ & Ritz$^a$        & Observed$^a$    &    Upper level                       &  Previous $^{a,c}$       & Ref.$^d$ \\  
wavelength    & wavelength       & wavenumber      & wavenumber      &                                      &  wavenumber             \\
     (nm)     &    (nm)          &  cm$^{-1}$      &  cm$^{-1}$      &                                      & (cm$^{-1}$)             \\
\hline                                                                                                                          
385.991112    &  386.100556 (16) & 25899.9886 (10) &         $^b$    & $\rm 3d^6(^5D) 4s4p(^3P)\, z^5D^{\circ}_4  $ &  25899.990  (5)  & (2) \\ 
371.993443    &  372.099252 (15) & 26874.5501 (11) &         $^b$    & $\rm 3d^6(^5D) 4s4p(^3P)\, z^5F^{\circ}_5   $ &  26874.551  (5)  & (2) \\ 
344.060549    &  344.159159 (14) & 29056.3238 (12) &         $^b$    & $\rm 3d^6(^5D) 4s4p(^3P)\, z^5P^{\circ}_3  $ &  29056.3240 (20) & (1) \\
302.063869    &  302.151846 (12) & 33095.9420 (13) &         $^b$    & $\rm 3d^7(^4F)4p\, y^5D^{\circ}_4          $ &  33095.943  (5)  & (2) \\ 
298.356948    &  298.444002 (12) & 33507.1234 (13) &         $^b$    & $\rm 3d^7(^4F)4p\, y^5D^{\circ}_3          $ &  33507.125  (5)  & (2) \\ 
296.689807    &  296.776446 (12) & 33695.3964 (14) & 33695.3957 (15) & $\rm 3d^7(^4F)4p\, y^5F^{\circ}_5          $ &  33695.399  (5)  & (2) \\   
271.902709    &  271.983269 (11) & 36766.9674 (15) & 36766.9670 (16) & $\rm 3d^6(^5D) 4s4p(^1P)\, y^5P^{\circ}_3  $ &  36766.9678 (20) & (1) \\  
252.284917    &  252.360810 (10) & 39625.8040 (16) & 39625.8026 (17) & $\rm 3d^6(^5D) 4s4p(^1P)\, x^5D^{\circ}_4  $ &  39625.804  (5)  & (2) \\                     
250.113178    &  250.188564 (10) & 39969.8524 (16) & 39969.8533 (17) & $\rm 3d^6(^5D) 4s4p(^1P)\, x^5D^{\circ}_3  $ &  39969.8522 (20) & (1) \\  
248.327097    &  248.402069 (10) & 40257.3136 (16) & 40257.3113 (21) & $\rm 3d^6(^5D) 4s4p(^1P)\, x^5F^{\circ}_5  $ &  40257.320  (10) & (2) \\                     
246.264702    &  246.339197 (10) & 40594.4328 (16) & 40594.4331 (17) & $\rm 3d^6(^5D) 4s4p(^1P)\, x^5F^{\circ}_4  $ &  40594.4319 (20) & (1) \\  
229.816879    &  229.887669 (9)  & 43499.5057 (17) & 43499.5059 (18) & $\rm 3d^6(^5D) 4s4p(^3P)\, w^5D^{\circ}_4  $ &  43499.5065 (20) & (1) \\  
216.677316    &  216.745313 (9)  & 46137.0991 (19) & 46137.1008 (19) & $\rm 3d^7(^4P)4p\, w^5P^{\circ}_3          $ &  46137.1012 (20) & (1) \\                           
\hline                                                                                                                                            
\end{tabular}                                                                                                                                                                        
                               
{\footnotesize
$^a$ One standard uncertainty in the last digits given in parenthesis.
$^b$ Line is self-reversed in spectra. Accurate wavenumbers cannot be determined.
$^c$ Wavenumber has been increased by 1.06 parts in 10$^7$ as recommended by 
\citet{Nave_11}.     
$^d$ (1) \citet{Nave_91}; (2) \citet{Nave_94}\\ 
}
\label{tab:FeI}                                                                                   
\end{table*}

\subsection{\ion{Ni}{ii}}

Eleven \ion{Ni}{ii} resonance lines have been previously identified as important for
studies of the time-variation of the fine structure constant by
\citet{Murphy_14}. Seven lines, measured by \citet{Shenstone1970} using grating
spectroscopy, are of insufficient accuracy for fine structure studies.
\citet{Pickering2000a} reported four lines, measured using high-resolution FT
spectroscopy of a nickel-neon HCL. The number and accuracy of
all eleven resonance lines can now be improved using measurements from the
comprehensive analysis recently completed by \citet{Clear2018}.

The new wavenumbers were measured in the spectrum of a nickel-helium HCL,
recorded using the Imperial College VUV-FT spectrometer \citep{Thorne_87}.
Observed and calculated Ritz wavenumbers for \ion{Ni}{ii} from this term
analysis are given in Table \ref{tab:NiII}. Lines without observed wavelengths
lie beyond the lower wavelength limit of the magnesium fluoride beam splitter
used in the Imperial instrument ($\approx$140 nm). Wavenumbers were calibrated
against selected Ar II standards \citep{Learner_88}, measured in the visible by
\citet{Whaling_95}, and the calibration propagated to the VUV by Ni lines in
overlapping spectral regions. The wavenumber uncertainty of the observed lines
is the addition in quadrature of the statistical uncertainty of the line
position and the calibration uncertainty of its spectrum. Ritz wavelengths were
calculated from energy levels optimised using the {\sc lopt} code of
\citet{Kramida_11}. A global calibration uncertainty, based on the uncertainty
of the Ar II standard lines, was estimated to be 4 parts in $10^8$ and was set
as the minimum uncertainty for all Ritz wavenumbers. Full details of the
\ion{Ni}{ii} spectra, calibration and term analysis are given in
\citet{Clear2018}.

Our new wavenumbers and wavelengths for these lines are presented in Table
\ref{tab:NiII_FTS_SNR}. The uncertainties are at least an order of magnitude
smaller than those previously measured using grating spectroscopy. The
uncertainty of lines previously measured in FT spectra have decreased by roughly
a factor of 2, due largely to the increase in signal-to-noise ratio of
\ion{Ni}{ii} lines measured in a HCL with helium as the carrier gas as opposed
to neon used by \citet{Pickering2000a}.
\begin{table*}                                                                                                        
        \caption{Ritz wavelengths and wavenumbers of resonance lines in \ion{Ni}{ii}}
                \begin{tabular}{llllll}
                        \hline
Ritz vacuum $^a$    & Ritz$^a$             & Observed$^a$   &    Upper level       &  Previous $^a$       & Ref.$^c$ \\
wavelength          & wavenumber           & wavenumber     &                      &  wavenumber          &      \\
(nm)                & (cm$^{-1}$)          & (cm$^{-1}$)    &                      &  (cm$^{-1}$)         &      \\
\hline                                     
175.191553(7)       & 57080.3778(23)       & 57080.3777(21) & $\rm 3d^8(^3F)4p\, ^2F_{7/2}   $ & 57080.377(4)$^b$  & (1)  \\
174.155319(7)       & 57420.0091(23)       & 57420.0100(21) & $\rm 3d^8(^3F)4p\, ^2D_{5/2}   $ & 57420.017(4)$^b$  & (1)  \\
170.960430(7)       & 58493.0678(23) & 58493.0671(21) & $\rm 3d^8(^3F)4p\, ^2F_{5/2}   $ & 58493.075(4)$^b$  & (1)  \\
170.341227(7)       & 58705.6944(23)       & 58705.695(5)   & $\rm 3d^8(^3F)4p\, ^2D_{3/2}   $ & 58705.711(15)$^b$ & (1)  \\
150.215259(6) & 66571.133(3)   & 66571.133(14)  & $\rm 3d^8(^3P)4p\, ^4P_{5/2}   $ & 66571.25(5)       & (2)  \\
146.776033(6)       & 68131.014(3)   & 68131.010(14)  & $\rm 3d^8(^1D)4p\, ^2F_{7/2}   $ & 68130.94(5)       & (2)  \\
146.726335(6)       & 68154.091(3)   & 68154.07(3)    & $\rm 3d^8(^1D)4p\, ^2D_{3/2}   $ & 68154.01(5)       & (2)  \\
145.484653(6)       & 68735.773(3)   & 68735.78(1)    & $\rm 3d^8(^1D)4p\, ^2D_{5/2}   $ & 68735.51(5)       & (2)  \\
139.332727(6)       & 71770.648(3)   & -              & $\rm 3d^8(^3P)4p\, ^2D_{5/2}   $ & 71770.51(5)       & (2)  \\
137.013652(5)       & 72985.428(3)   & -              & $\rm 3d^8(^3P)4p\, ^2P_{3/2}   $ & 72985.46(5)       & (2)  \\
131.722053(5)       & 75917.432(3)   & -              & $\rm 3d^8(^1G)4p\, ^2F_{7/2}   $ & 75917.46(5)       & (2)  \\
\hline
\end{tabular}

        $^a${\footnotesize One standard uncertainty in the last digits given in parenthesis.}
        $^b${\footnotesize Wavenumber has been increased by 6.7 parts in 10$^8$ as recommended by  \citet{Nave_11}.}
        $^c${\footnotesize (1) \citet{Pickering2000a}; (2) \citet{Shenstone1970}\\
        \label{tab:NiII}   }
\end{table*}

\begin{table}
  \centering
  \caption{Signal to noise ratios of observed \ion{Ni}{ii} lines in FT spectra}
     \begin{tabular}{llll}
        \hline
        Wavenumber  & Wavelength & SNR$^a$   & SNR $^b$  \\
        (cm$^{-1}$) &   (nm)     &  &  \\
        \hline                   
        57080.3778  & 175.191553 &  224  &  45  \\
        57420.0091  & 174.155319 &  273  &  60  \\
        58493.0677  & 170.960430 &  163  &  55  \\
        58705.6944  & 170.341227 &  23   &  9   \\
        \hline
     \end{tabular}
     
     {\footnotesize
     $^a$  SNR in \citet{Clear2018};
$^b$ SNR in \citet{Pickering2000a}
}

        \label{tab:NiII_FTS_SNR}
\end{table}

\section{Conclusions}                                                         

We have measured improved wavelengths for lines of \ion{Si}{ii}, \ion{C}{ii},
\ion{Fe}{i}, and \ion{Ni}{ii} of importance for detecting possible changes in
the fine structure constant, $\alpha$, in the early Universe. The majority of
the Ritz wavelengths have been derived from energy levels optimized using FT
spectroscopy. The uncertainty of the Ritz wavelengths ranges from
$2\times10^{-6}$~nm to $3\times10^{-4}$~nm, depending on the wavelength region.
Comparison of the wavelength scales of the \ion{Si}{ii} lines with
KE74 shows that while their wavelengths agree within the joint
uncertainties, better agreement is achieved by reducing their wavelengths by
5$\times$10$^{-5}$~nm. This was used to derive Ritz wavelengths and wavenumbers
for an additional six lines of \ion{Si}{ii} using their wavelengths. Wavelengths
of three \ion{C}{ii} lines near 135~nm have been measured in one FT spectrum
that confirm the previous measurements of \citet{Herzberg_58}. The uncertainties 
of our new Ritz wavelengths for \ion{Fe}{i} and \ion{Ni}{ii} are over a factor 
of 2 lower than previous wavelengths measured using FT spectroscopy, and over a 
factor of 10 lower than those measured using grating spectroscopy.

\section*{Acknowledgements}
This work was partly supported by NASA  under interagency agreements NNH14AY78I 
and NNH17AE081, and the STFC of the UK.                                                        

\section*{Data Availability}
The data underlying this article will be shared on reasonable request 
to the corresponding author.

\bibliographystyle{mnras} 
\bibliography{alpha_13Aug20}  
       
\label{lastpage}
\end{document}